# Dynamic System Adaptation by Constraint Orchestration

L.P.J. Groenewegen[1,⋆] and E.P. de Vink[2]

[1] FaST Group, LIACS, Leiden University
[2] Formal Methods Group, Eindhoven University of Technology

**Abstract.** For Paradigm models, evolution is just-in-time specified coordination conducted by a special reusable component McPal. Evolution can be treated consistently and on-the-fly through Paradigm's constraint orchestration, also for originally unforeseen evolution. UML-like diagrams visually supplement such migration, as is illustrated for the case of a critical section solution evolving into a pipeline architecture.

## 1 Problem Situation

Software systems are large and complex. However, more strikingly, software systems have a strong tendency to grow over time, both in size and complexity. In order to deal with size and complexity, software architectures are used. A software architecture provides a global description of an actually far more detailed software system by giving an overview in terms of *components* and *links*. Components are the main relevant parts, links are the relevant connections between them.

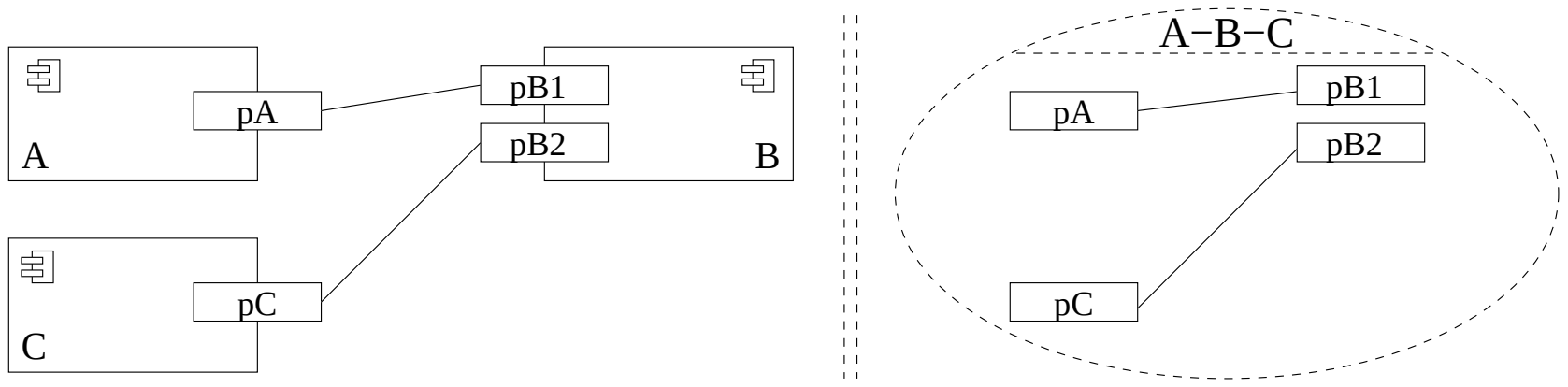

**Fig. 1.** Two composite structure diagrams.

Figure 1 presents two architectural visualizations in UML-style [14]. The more usual, fully structural style is at the left. The larger, iconized rectangles are *components*. The smaller rectangles, positioned across a components border, are *ports*, each one representing an interface offered by that component. A port serves as the scene of action for a *role* that the component is responsible for in the architectural constellation. Lines connecting ports are *links*, via which components communicate by executing their roles. At the right of Figure 1, a less common, more dynamically oriented presentation is given. It visualizes a *collaboration*: a grouping of roles constituting a cooperative unit, a protocol. The roles are represented via their respective ports; components remain hidden.

⋆ Corresponding author, e-mail luuk@liacs.nl.

An architecture serves as a basis for the global understanding of the system in terms of coherence between its constituents: components, ports/roles, links and collaborations/protocols. The coherence covers the structural fitting of these four constituent categories, and, more importantly, also their dynamic consistency. In particular, each category of constituents has its own type of dynamics: A) For a component it is local, internal component behaviour, usually hidden. B) For a port it is local, external visible role behaviour. C) For a link it consists of sending and receiving in either direction, role interaction. D) For a collaboration it is the coordination of the role behaviours and their interactions into an overall protocol. We refer to these four types of dynamics as type A, B, C and D, respectively.

In view of the mutual dynamic fitting of behavioural specifications, coherence between such specifications is of utmost relevance. In Figure 2, three situations are being indicated, T1–T3, where coherence of dynamics has to be assured. In line with [21], we call such a situation a dynamic consistency problem type.

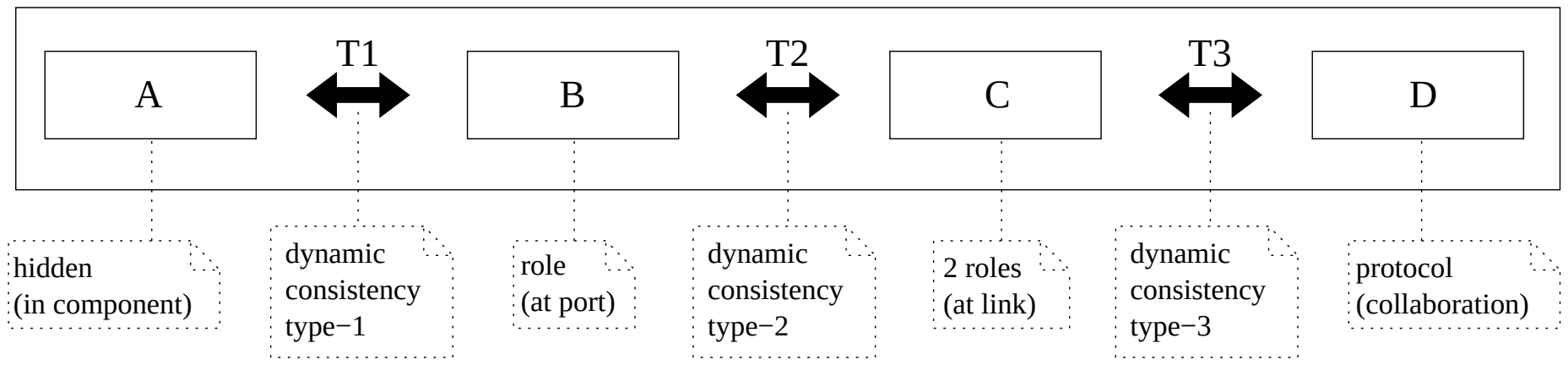

**Fig. 2.** Dynamic consistency problem types T1 to T3 in a stable UML collaboration.

In addition to the dynamics and consistency relevant for the foreseen situation of regular, stable collaboration, there may be evolution from the original, foreseen situation towards the initially unforeseen situation of a renewed collaboration. During migration in particular, i.e. when taking an evolutionary step, behavioural specifications, of any type A–D, can change. Figure 3 visualizes four additional dynamic consistency problem types, T4–T7, resulting from migration. Generally, to maintain consistency, the change of dynamics should occur smoothly: consistent with what preceded as well as consistent with what is to come next. So, behavioural execution of all dynamics should be deflected sufficiently gradually. A primary question then is, how can the dynamics of components, ports, links and collaborations be represented, to support the claim that the system behaviour remains coherent during the complete migration. Indeed, such coherent execution, solving dynamics consistency problem types T1–T3, should not only last during the original, stable situation, but also during the migration, as well as during the new situation, uninterruptedly that is, be it changing gradually. Here, as is typically the case for large software systems, executions are assumed to continue, without any halting or other form of quiescence.

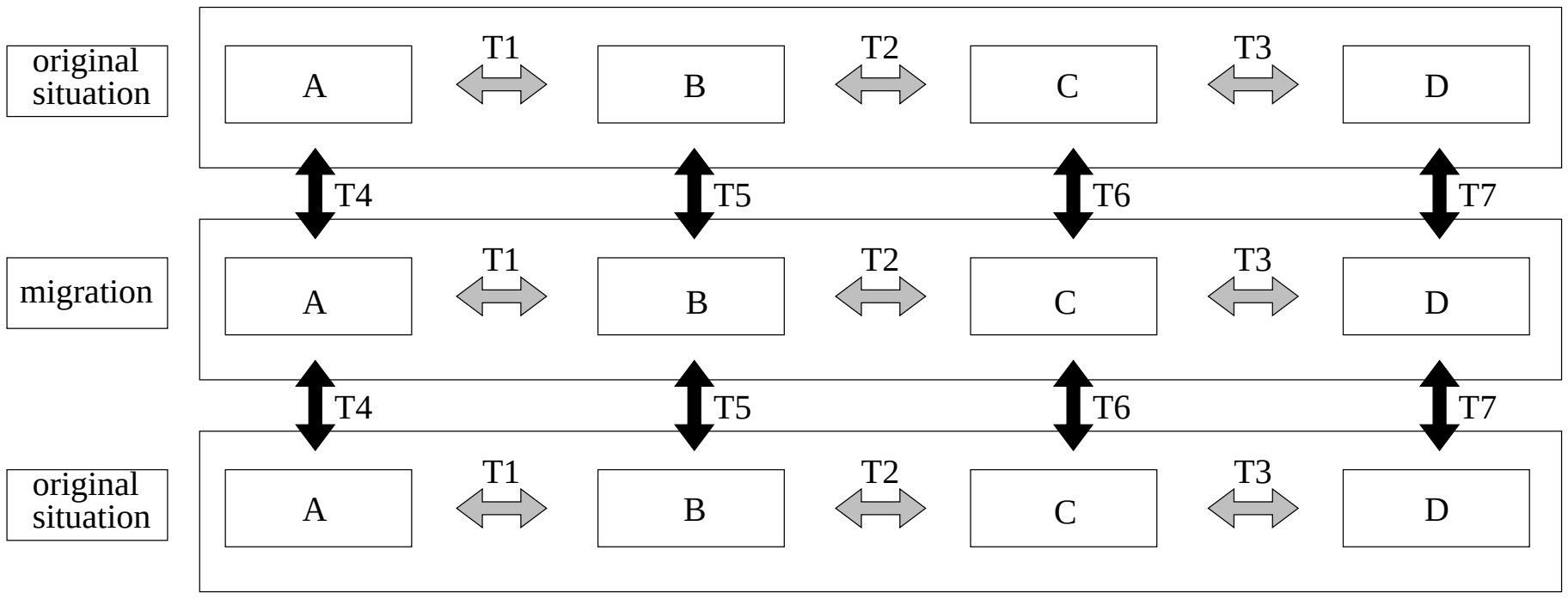

**Fig. 3.** Dynamic consistency problem types T4 to T7 in case of migration.

As a solution to the above consistency problems, we propose the coordination modeling language Paradigm, see [15, 17], together with the special component called McPal (short for Managing changing Processes ad libitum).[3] A Paradigm model without McPal specifies coordination of stable, foreseen collaborations by dynamically restricting and re-enabling parts of component behaviour: constraint orchestration. In this manner, a Paradigm model provides coherence of type T1–T3 between dynamics of type A, B, C and D. In view of future, unforeseen migration of a Paradigm model as-is, McPal is to be incorporated into it. Component McPal, on the basis of the Paradigm notions of interaction, is designed as follows. First, McPal allows for extending the constraints and their dynamic compositions while keeping the execution of the system going as-is. Second, McPal coordinates migration from the as-is execution to the to-be execution aimed at on the basis of the new constraints and their new orchestration recently added. Third and last, once the execution situation aimed at has been established, McPal removes behaviour, constraints and orchestrations no longer needed. Note, executions are assumed to continue, without any halting or quiescence.

In the field of software migration and evolution, the present work fits in the setting of unanticipated dynamic software evolution, focusing on processes and their dynamics. An example thereof is dynamic co-evolution, where changing business rules and evolving software need to be aligned. A number of general techniques are proposed in [23, 24]: McPal's migration control realizes their generation technique; complementarily, Paradigm, via its phases, traps and consistent dynamic character, covers their other techniques: decomposition (uncom-monly dynamic via processes and their phases), reification, reflection (both via consistency) and probes (via traps).

[3] For a restricted version of McPal of a fixed form, see [16].

In the setting of component-based software engineering, process languages and mobile calculi are exploited to formally analyze runtime modification of adaptors, for example via context-dependent dynamic mappings. Cf. [6, 9, 8, 20]. Also, for the non-dynamic as well as the dynamic case, algorithmic procedures to derive concrete adaptors from high-level specification are proposed [5, 2, 26]. Reconfiguration, mainly at the structural level, is treated by graph grammars and graph transformation, e.g., in [13, 19, 3, 22, 4]. A recent development based on hierarchical rewriting can be found in [7].

Translation of UML models via Object-Z into CSP, has been proven sound for so-called basic consistency [27], whereas verification of UML sequence diagrams are discussed in [28]. The paper [10] studies behaviour preserving model transformations composed from subtransformations per view and provide a prooftechnique to achieve global correctness. The views are similar to our collaborations. However, in our approach Paradigm guarantees global correctness of evolution by construction. Architectural adaptation by graph transformations, as an implicit McPal in our terms, has been studied by various authors, [25] being reminiscent to our approach. Consistency in the context of evolution and self-management is also addressed, e.g., in [12, 11, 29].

The remainder of the paper is structured as follows: In Section 2, Paradigm's way of modeling is illustrated for a critical section example with known dynamics. In Section 3 it is discussed how McPal deals with unforeseen behaviour, with the example migrating to a producer-consumer pipeline; additional UML diagrams prove useful too. Section 4 gives some concluding remarks. The appendix presents an overview of Paradigm notions.

## 2 Constraint Orchestration for Foreseen Coordination

This section briefly introduces the coordination modeling language Paradigm in terms of two kinds of behavioural constraints, *phase* and *trap*, and two ways of dynamically composing these constraints, sequentially as *global process* and synchronized as *consistency rule*. A more detailed description of Paradigm's core definitions can be found in the appendix.

A Paradigm process is given by a state transition diagram (STD), its transitions labeled by actions. Usually a process is visualized by a simple statemachine diagram. In its detailed form, a process captures type A dynamics, as discussed in Section 1. See the middle part of Figure 4 for two example processes (to be explained later): a worker and a scheduler.

A *phase* is a restricted version of an underlying process, obtained by constraining to a part of the process behaviours. A *trap* is a further constraint on it: a trap is a subset of the phase states, which cannot be left by means of the transitions available during the phase. A trap is committed to autonomously, just by entering it; a phase is imposed, from elsewhere. The lower-left part of Figure 4 depicts three phases of the worker, *OutCS*, *Inspec* and *InCS*, and four traps, *triv*, *notyet*, *started* and *done*.

Given some phases and traps of a process, the induced global process has the phases as its states. The global process has a transition from one phase to another, if there is a trap of the first phase that is contained in the second phase. So, a state in the trap is a state of the first phase as well as a state of the second phase. The trap is referred to as a connecting trap and is used as the label of the transition of the global process. The lower-right part of Figure 4 shows an example of a global process, induced by the phases and traps from the lower-left figure part.

In a sense, a global process is a sequential composition of phases. It specifies a role, at a port, of the component whose hidden dynamics is the underlying process. It specifies Section 1's type B behaviour. Moreover, a global process brings forward type C dynamics: Information about a trap committed to *is sent from* the role's port via the link. Information about a phase imposed, renewed constraint for the process behaviour regarding the role, *is received* at the same role's port where the type B behaviour is occurring. A mirrored version of the role is occurring at the link's other end, modulo possible delays between sending from one end and receiving at the other; time-ordered sends/receives constitute type C behaviour.

Given mirrored roles, a *consistency rule* synchronizes their steps. It couples each synchronized step to a step of a detailed process, a so-called manager process. The possible sequences of the synchronized steps constitute a protocol, Section 1's type D dynamics. Thus, roles can be seen as protocol roles. The protocol constitutes the coordination of the collaboration by properly composing the relevant constraints. We call this *constraint orchestration.* On the basis of Paradigm's notions, the dynamic consistency problem types of the stable, foreseen coordination situation are made explicit. A UML-like visualization of this coordination stresses the coherence. Figure 4 presents a small example of a so-called critical section management problem (CSM), with its structure diagrams and with various fragments of the Paradigm model, including detailed processes, phases, traps and global processes.

Figure 4's upper layer gives the collaborating participants of the CSM architecture: Worker components competing for permission to enter their critical section. A scheduler component, for three workers, giving the permission to a worker asking for it, exclusively; the permission is withdrawn only after the worker indicates it does not need it any longer. The middle layer of Figure 4 gives the processes for workers and scheduler. A worker needs the permission for being in its state *Crit* where it does its critical activity. In *Post* it does some wrapping up, in *Free* is does nothing in particular, in *NCrit* it does non-critical work and in *Pre* it prepares its critical activity. As long as it does not have the permission to go to *Crit*, it waits there. In UML-style, the black dot with outgoing edge indicates the starting state *Free*. Likewise, process *Scheduler* starts in $Check_1$. In a state $Check_i$, the scheduler checks whether $Worker_i$ wants to have the permission. If so, it goes to $Asg_i$ where it assigns the permission to $Worker_i$;

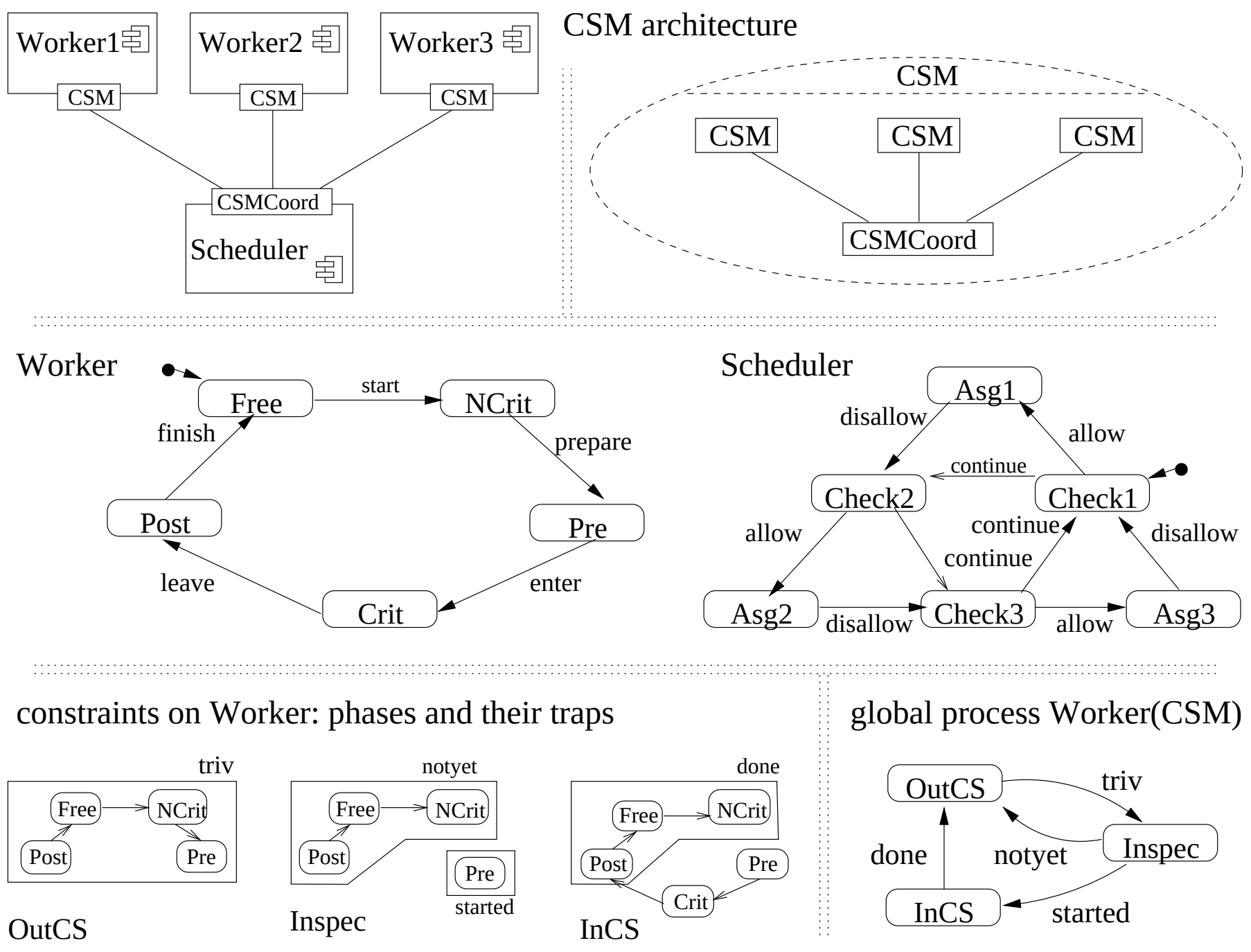

**Fig. 4.** CSM collaboration.

if not so, it goes to the next state $Check_{i+1}$ in round robin fashion. In $Asg_i$ it waits until $Worker_i$ has finished its critical activity.

Figure 4's lowest layer visualizes constraints, viz. three phases *OutCS*, *Inspec* and *InCS*, each as partial STD of a worker, and four traps, *triv* of *OutCS*, *notyet* and *started* both of *Inspec* and *done* of *InCS*, each drawn as polygon around the states it consists of. Being a commit within the phase, a trap once entered cannot be left as long as the phase constraint holds. The phases and traps reflect the following intuition. Phase *OutCS* covers the behavioural phase where a worker does not have the permission to enter its critical section. *OutCS* reflects, it is as if state *Crit* does not exists. Contrarily, *InCS* covers having that permission. *InCS* reflects, state *Crit* can be entered, but once only, as state *Pre* is unreachable during this behavioural phase. Phase *Inspec* is an interrupted form of *OutCS*, as entering state *Pre* is no longer possible during it: either trap *started* has been entered or trap *notyet*. Being in *started* is taken as a CSM-permission request, being in *notyet* is taken as no CSM-permission request yet. The trivial trap *triv* of *OutCS* reflects, *OutCS* can be interrupted –towards phase *Inspec*– unconditionally; trap *done* of *InCS* reflects, the CSM-permission can be withdrawn: state *Crit* has been left. At the right of Figure 4's lowest level, the global process or role $Worker_i(CSM)$ is given.

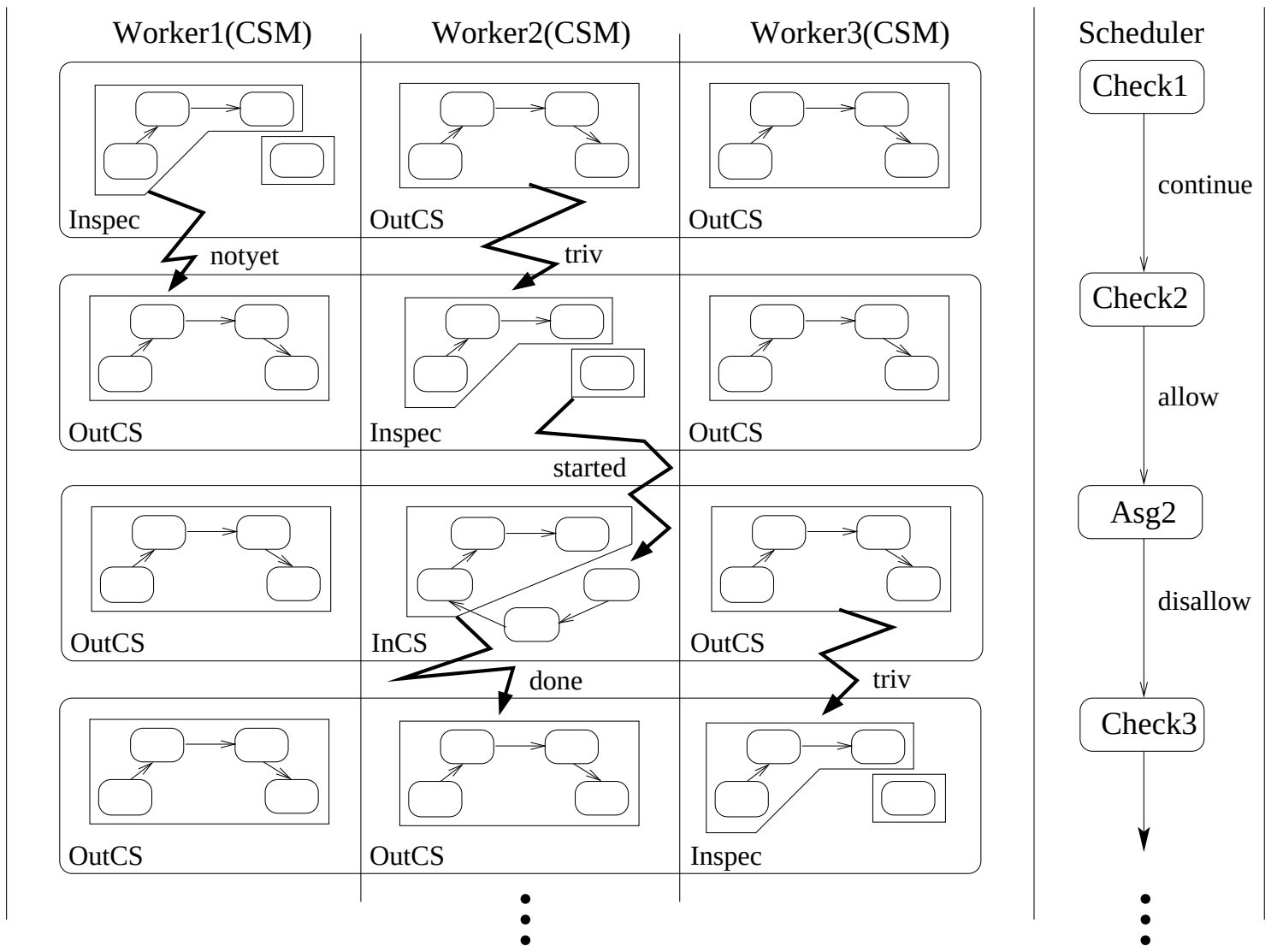

**Fig. 5.** CSM protocol: phase constraints enforced, trap constraints checked.

Synchronized realizations of the global behaviours of three workers constitute a protocol realization or scenario. Figure 5 gives such a scenario in the form of a UML-like activity diagram: three swimlanes for the three global processes coupled to one swimlane for the scheduler. In particular, our visualization reveals the behavioural consequences of the various phase constraints for the detailed behaviours after each protocol step. One sees, how, when and why exactly the CSM-permission is given exclusively to one worker at a time, indeed in round robin order, the phase *Inspec* shifting from $Worker_1$, to $Worker_2$, to $Worker_3$. The protocol steps as visualized in Figure 5, are specified by consistency rules, synchronizing global process steps and coupling them to one detailed manager step, as follows.

$$Scheduler\colon Check_i \stackrel{allow}{\to} Asg_i \ * \ Worker_i(CSM)\colon Inspec \stackrel{started}{\to} InCS \tag{1}$$

$$\begin{aligned} &Scheduler\colon Asg_i \stackrel{disallow}{\to} Check_{i+1} \ * \\ &\quad Worker_i(CSM)\colon InCS \stackrel{done}{\to} OutCS,\ Worker_{i+1}(CSM)\colon OutCS \stackrel{triv}{\to} Inspec \end{aligned} \tag{2}$$

$$\begin{aligned} &Scheduler\colon Check_i \stackrel{continue}{\to} Check_{i+1} \ * \\ &\quad Worker_i(CSM)\colon Inspec \stackrel{notYet}{\to} OutCS,\ Worker_{i+1}(CSM)\colon OutCS \stackrel{triv}{\to} Inspec \end{aligned} \tag{3}$$

In fact, Figure 5's first step illustrates rule (3): a scheduler transition from a *Check* state to a next *Check* state is coupled to two global process transitions: one for global process $Worker_i(CSM)$, returning from being interrupted in *Inspec* to not having the permission in *OutCS* as trap *notyet* was entered; the other for the next global process $Worker_{i+1}(CSM)$, changing from not having the permission in *OutCS* to being interrupted in *Inspec* as trap *triv* was entered trivially. Similarly, Figure 5's second step illustrates rule (1) and Figure 5's third step illustrates rule (2).

In order for a rule to be applied, Paradigm's definitions in addition require: the one detailed transition mentioned in the rule, occurs in every currently valid phase constraint as specified by the various current states of the relevant global processes. Based on this, Paradigm models for foreseen coordination succeed in guaranteeing dynamic consistency of type T1 to type T3.

In the critical section management example above, the scheduler is also referred to as the *manager* of the collaborations, the workers as its *employees*. As such, the scheduler is occurring at the left-hand side of a consistency rule, the workers on the right-hand side. In the collaboration *CSM*, the local STD behaviour of the manager, i.e. the *Scheduler* process is type A dynamics, the role behaviour of the global processes $Worker_i(CSM)$, the employees, is type B dynamics. The trap information from the workers as employees to the scheduler as manager and, vice versa, the newly prescribed phases flowing from manager to employees, constitute type C dynamics. The total of all consistency rules together form the protocol, type D dynamics, comprising the coordination of the components in the collaboration.

## 3 Evolution by Constraint Orchestration

In view of unforeseen change within an architecture, the special component *McPal* is to be added to it: for coordinating unanticipated migration towards a new way of working. During a stable collaboration phase, *McPal* is stand-by only, not influencing the other components at all. But by being there, *McPal* provides a pattern for dynamic evolution management in the architectural constellation of the Paradigm model. To that aim, ports and links are in place, realizing rudimentary interfacing for the moment. As soon as, via *McPal* but in strict separation of the model's stable working, a new way of working together with a migration towards it, has been modeled, typically just-in-time (JIT), *McPal* starts coordinating the migration. Its own migration begins, the migration of the others is started thereafter. Finishing the migration is done in reversed order. The others are explicitly left to their new stable collaborative work before *McPal* ceases to influence the others. It is stressed, that the migration is on-the-fly. *McPal* activities and migration to the new evolution phase can be mixed with older behaviour.

Figure 6 visualizes *McPal*'s hidden, detailed dynamics (type A) as follows. In its starting state *Observing*, *McPal* is doing nothing in particular, but it can observe, that something should change. State *JITting* is where just-in-time foreseeing and modeling of such a concrete change occurs. The extended model then is available in state *NewRuleSet*. So, upon leaving that state for state *StartMigr*, *McPal* is supposed to change its own phase *StatPhase* into an originally unknown next phase *MigrPhase*, which by then is known indeed. See Figure 6 for these two phases with their relevant traps and for the induced global process *McPal*(*Evol*).

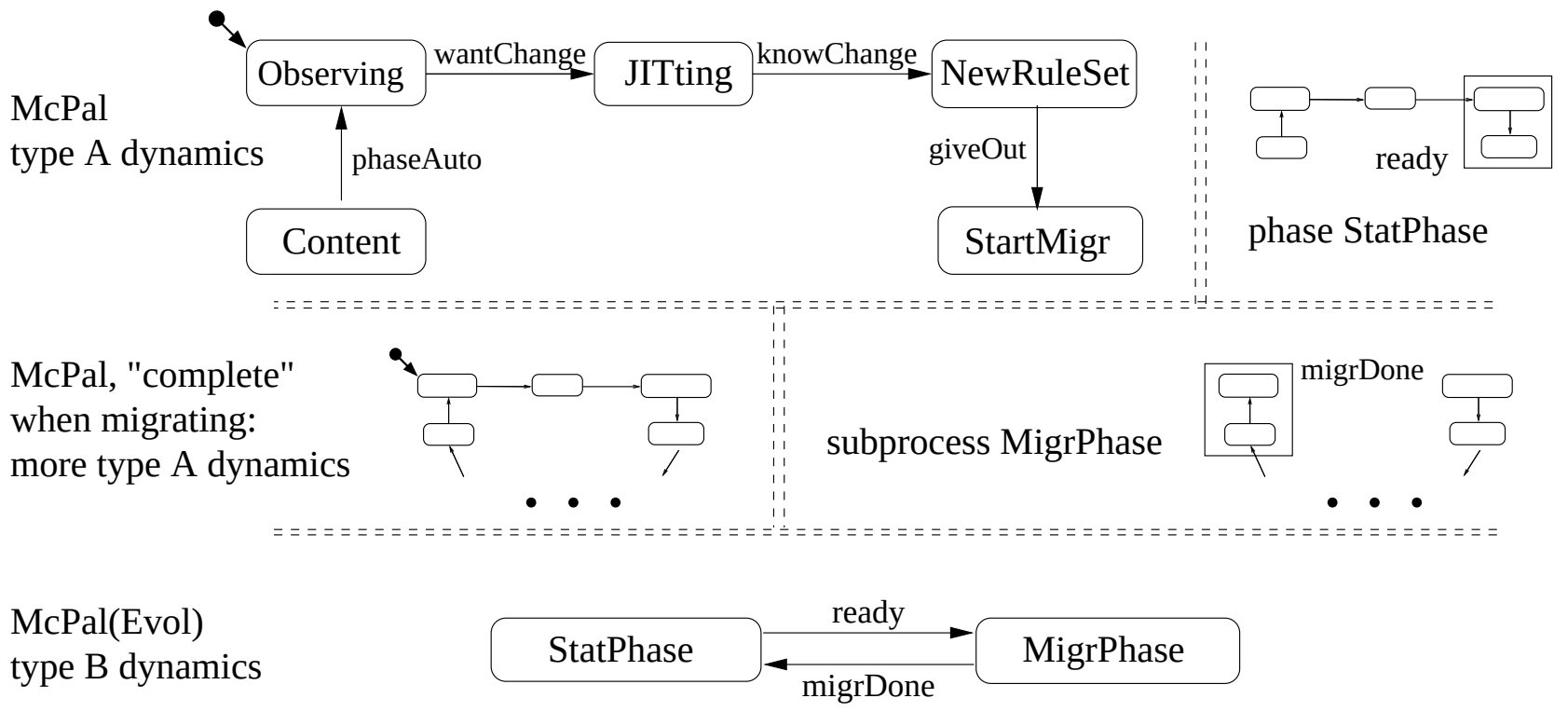

**Fig. 6.** *McPal* - during a stable collaboration situation 'mainly'.

Figure 7 gives an overview of all components cooperating in collaboration *Evol*. *McPal* has the *EvolCoord* role, which here means, it is manager of five employees having an *Evol* port each: three workers, a scheduler and *itself*. New constraints are imposed on the employees according to the migration details.

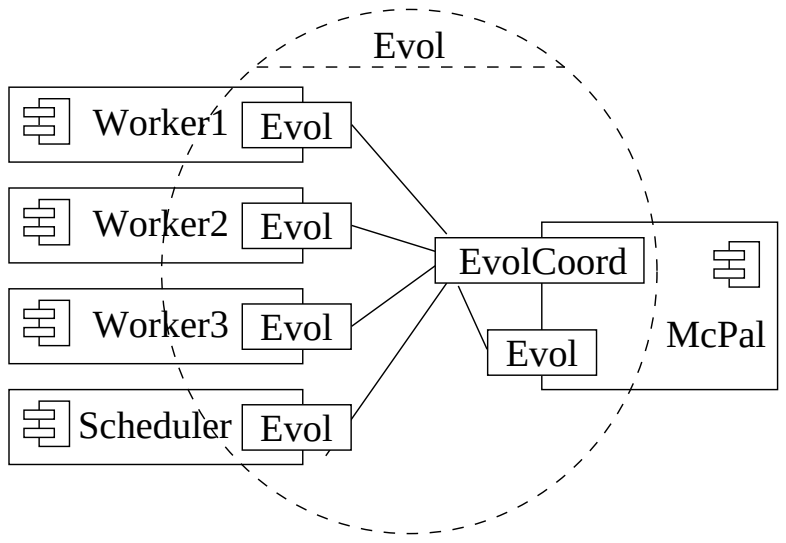

**Fig. 7.** The other components.

Figure 8 depicts a very small part of the type D dynamics of collaboration *Evol*, viz. the constraint orchestration fragment most essential for the *Evol* role of *McPal*. Similar as before, it is built by synchronizing all five *Evol* roles of the respective components coupled to the detailed steps of manager *McPal*. The Paradigm model for *McPal* has initially the following two consistency rules specifying the semantics for *McPal*'s first two steps.

$$McPal\colon Observing \stackrel{wantChange}{\rightarrow} JITting$$
$$McPal\colon JITting \stackrel{knowChange}{\rightarrow} NewRuleSet \,*\, McPal\,[\,Crs := Crs + Crs_{migr} + Crs_{toBe}\,]$$

The first step from state *Observing* to *JITting* has no coupling, so Figure 8 does not show a corresponding role step. The second step from *JITting* to *NewRuleSet* has no coupling either, but here, via a so-called change clause, the set of consistency rules *Crs* for the stable original situation is extended with the rules $Crs_{migr}$ for the migration only and with the rules $Crs_{toBe}$ for the new, stable situation to migrate to. This means in particular, apart from all other migration coordination details, *McPal* from then on has at least two more consistency rules:

$$McPal\colon NewRuleSet \stackrel{giveOut}{\rightarrow} StartMigr \,*\, McPal(\,Evol)\colon StatPhase \stackrel{ready}{\rightarrow} MigrPhase$$

$$McPal\colon Content \stackrel{phaseAuto}{\rightarrow} Observing \,*\, McPal(\,Evol)\colon MigrPhase \stackrel{migrDone}{\rightarrow} StatPhase,\ McPal\,[\,Crs := Crs_{toBe}\,]$$

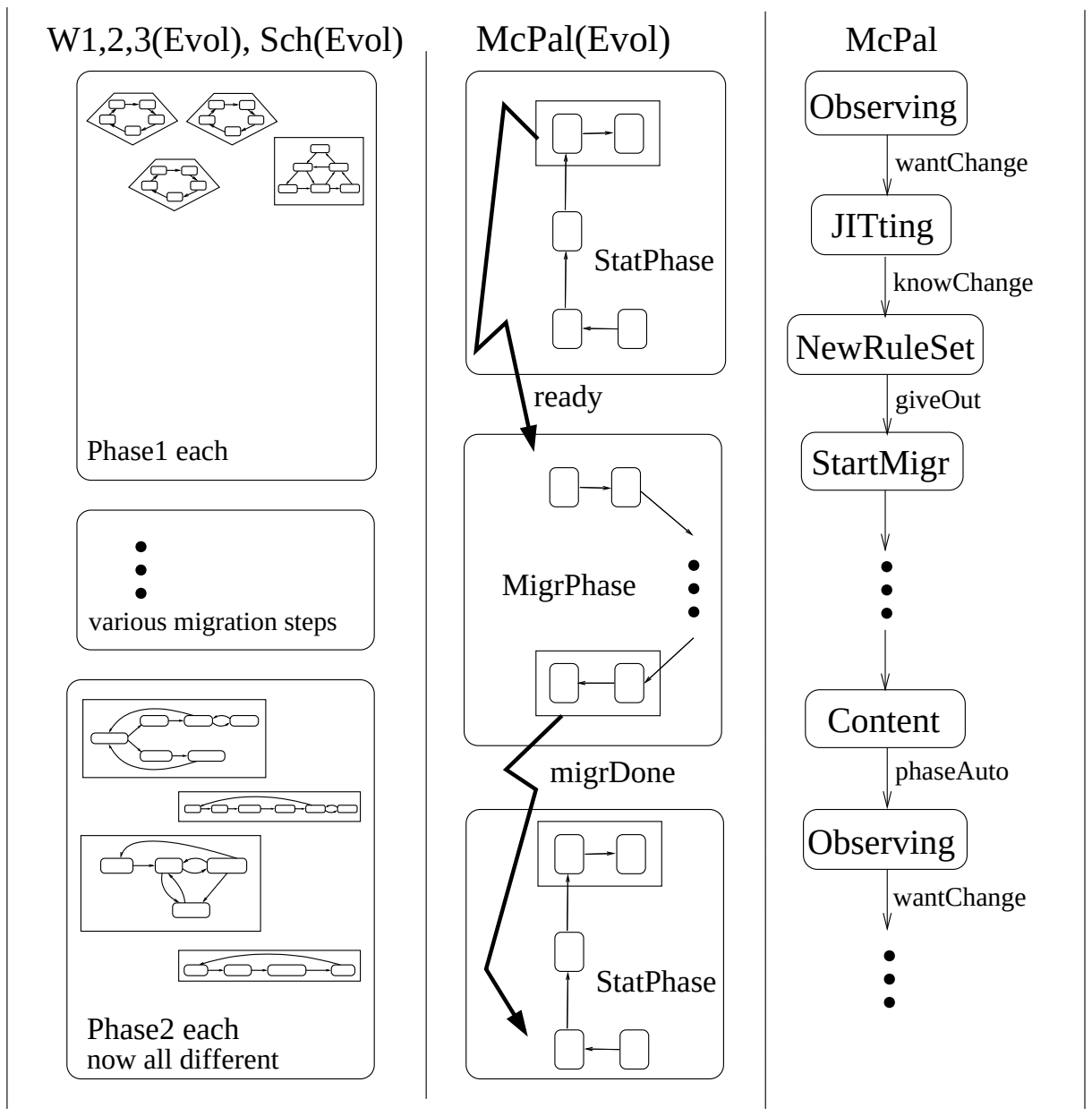

**Fig. 8.** Migration coordination as constraint orchestration.

The first new rule says, on the basis of having entered trap *ready*, the phase change from *StatPhase* to *MigrPhase* can be made, coupled to *McPal*'s transition from state *NewRuleSet* to *StartMigr*. Figure 8 expresses this where the upper 'lightning' step draws the reader's attention. One clearly sees, the three workers and scheduler remain the same, as there is no constraint change for them yet. From then on, the migration is a matter of normal coordination only, exactly according the planning as modeled while in state *JITting*. Eventually, *Scheduler* and each $Worker_i$ are restricted to new phases (that remain unexplained here). Moreover, their new coordination has by then been adapted as planned. So, consistency is accounted for by normal Paradigm execution. At the last migration step, after having phased out the dynamics that is no longer needed for the other components, i.e. after having entered trap *migrDone* of its phase *MigrPhase*, *McPal* returns from *MigrPhase* to *StatPhase* by making the (coupled) step from state *Content* to *Observing*. Then also the rule set *Crs* is reduced to $Crs_{toBe}$, by means of a suitable change clause. This can be seen in Figure 8 (lower 'lightning') and is expressed by the corresponding new rule for *McPal*. Once returned in state *Observing*, *McPal* is stand-by again, ready for a next migration.

In order to illustrate the general usability of *McPal* for migration coordination, we present the following example. Assume, *McPal* during its sojourn in state *Observing* wants to change the hierarchical architecture of the scheduler and its three worker processes into a pipeline of four processes, rather different in behaviour each. Assume in addition, in its state *JITting* process *McPal* establishes a Paradigm model for this particular situation to-be, as well as for a sufficiently smooth migration from the old, still current situation to the situation to-be.

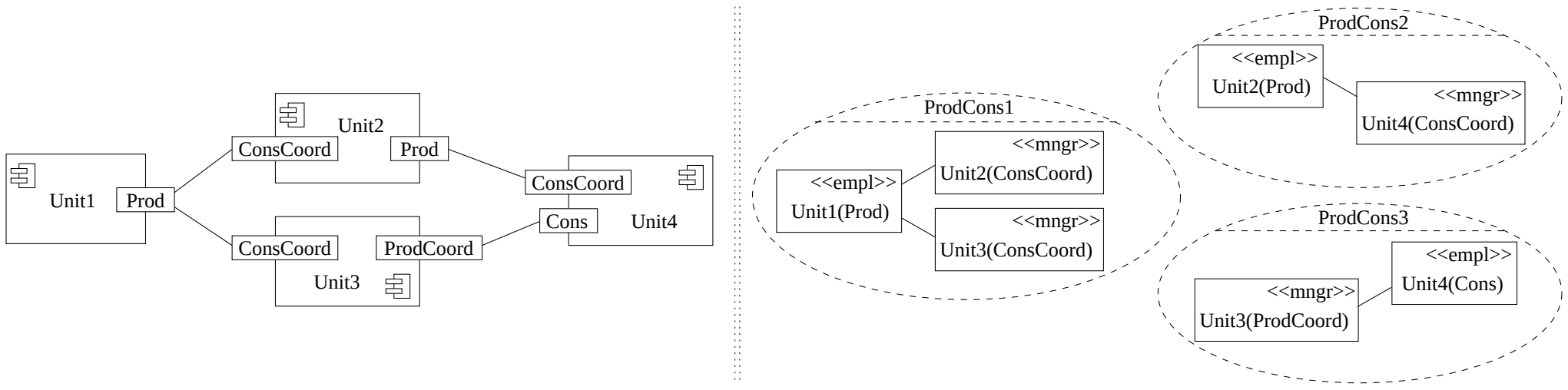

**Fig. 9.** New architecture and new collaborations.

Figure 9 presents our example to-be architecture of the pipeline: some item handling is repeatedly started by $Unit_1$, it is continued by either $Unit_2$ or $Unit_3$, and in both cases it is finished by $Unit_4$. Instead of the one collaboration *CSM* in the original architecture, now there are three collaborations of a producer-consumer character each: in $ProdCons_1$, employee process $Unit_1$ produces for its two consuming manager processes $Unit_2$ and $Unit_3$; in $ProdCons_2$, employee process $Unit_2$ produces for the consuming manager process $Unit_4$; in $ProdCons_3$, manager process $Unit_3$ produces for the consuming employee process $Unit_4$. Note, the three processes $Unit_1$, $Unit_2$ and $Unit_4$, constituting the upper part of the pipeline, have one employee role each, so they have each one global process; process $Unit_3$ has no employee role whatsoever, so it does not have a global process.

To illustrate the migration from the original *CSM* architecture to the new pipeline architecture, it is already very clarifying to guess how the original *CSM* collaboration is shrinking gradually (to nothing), while the other three new collaborations *ProdCons* are growing gradually (from nothing). One goal thereby is, for the sake of the example, that the original scheduler process has to become the new $Unit_3$ process – being the only process without employee roles, i.e. having no global process, both in the original situation and in the to-be situation.

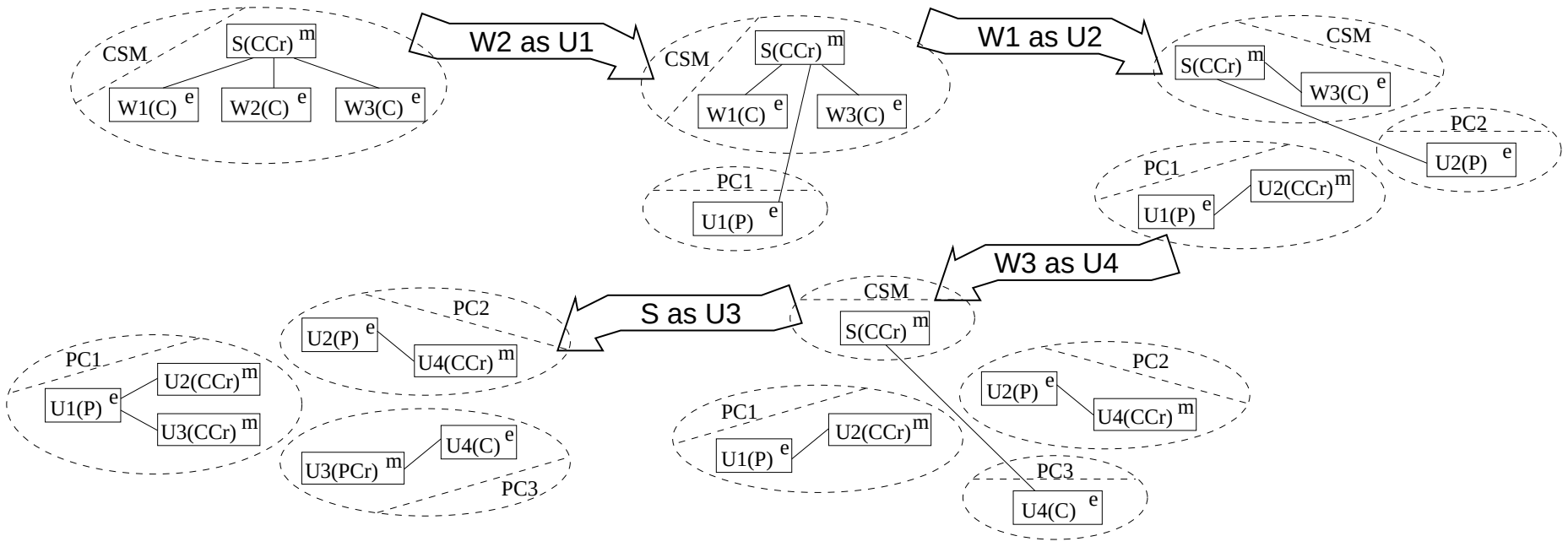

**Fig. 10.** An example scenario for migrating collaborations in snapshots.

Furthermore, each worker process may become any of the processes $Unit_1$, $Unit_2$ and $Unit_4$.

Figure 10 presents an example collaboration migration trajectory in three intermediate snapshots. Note how first the pipeline's upper part is being built from left to right. So, a first worker ($W_2$), upon becoming available, is to change into $Unit_1$, a second worker ($W_1$), upon becoming available, is to become $Unit_2$ and the remaining worker ($W_3$) is to become $Unit_4$, but not before this one has become available too. As long as a worker is not available yet, process *Scheduler* might be needed for regulating the critical section access. So, *Scheduler* is the last process that is to change into the last *Unit* process, $Unit_3$. Note, six different migration trajectories exist for this migration: $3 \times 2 \times 1$, as there are three possibilities for being the first available Worker, two possibilities for being the second available Worker and no choice for the last available Worker.

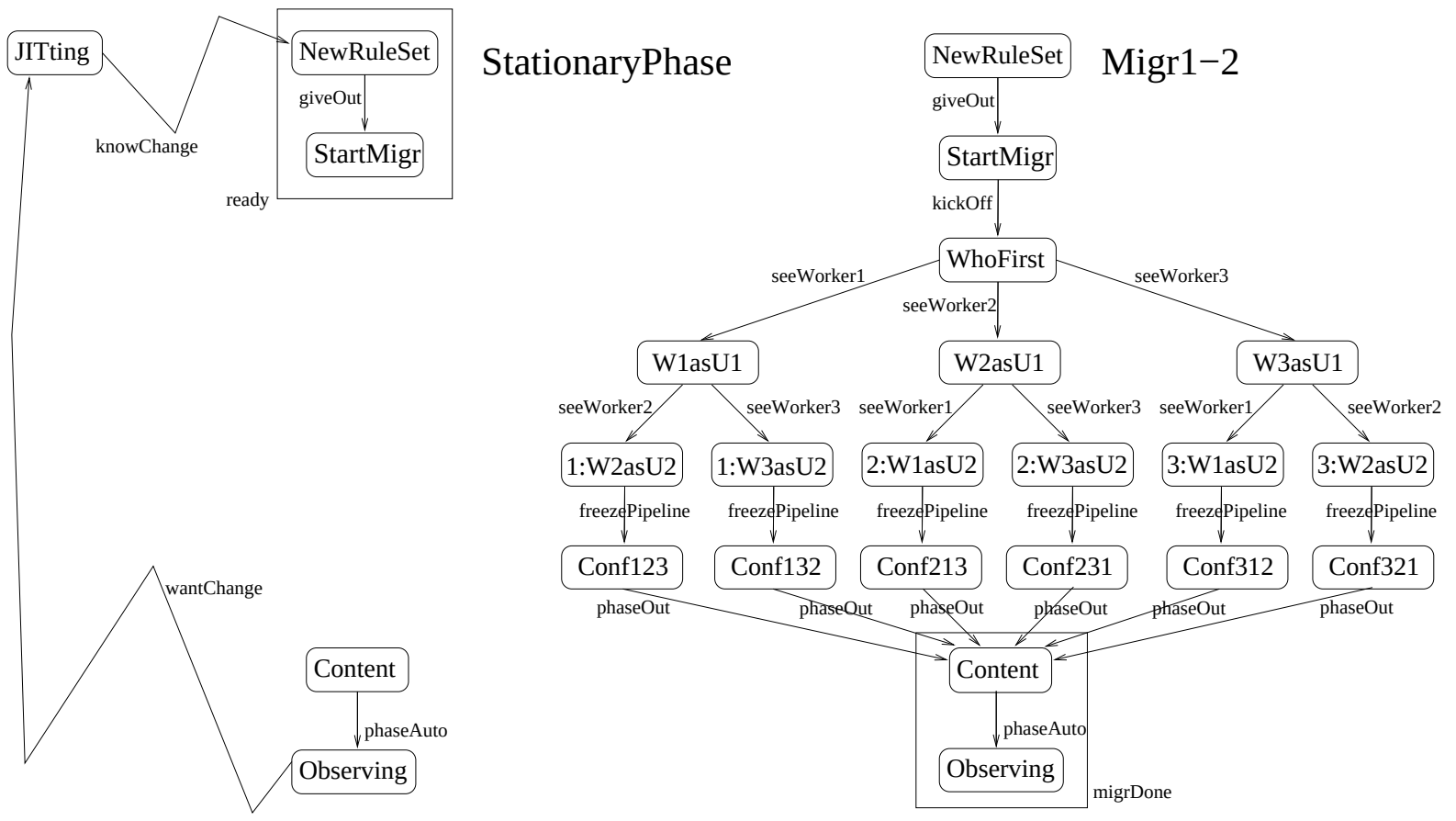

**Fig. 11.** *McPal* as migration coordinator.

The above example trajectory in snapshots presents a guideline for the migration coordination to be conducted by *McPal*. Figure 11 presents such coordination as *McPal*'s behaviour belonging to its phase $Migr_{1-2}$. Note, six different behaviours are possible, representing all different migration trajectories. In state *WhoFirst* process *McPal* waits until a first worker becomes available. In state $W_i\,\mathtt{as}\,U_1$ process *McPal* has found $Worker_i$ available first. In a state $i{:}W_j\,\mathtt{as}\,U_2$ it has found $Worker_j$ available next. The remaining Worker is $Worker_k$. In state $Conf_{i,j,k}$, it has found $Worker_k$ available and *Scheduler* has started the last turn it can give. So, after that turn *Scheduler* can become $Unit_3$. In state *Content* it has found, the scheduler and all workers have started to behave as the right unit. Finally, in state *Observing* it has swapped back from phase $Migr_{1-2}$ to *StatPhase*. So, then this migration is history.

Without presenting the details of the global processes at the *Evol* ports, we nevertheless sketchily visualize the subsequently valid behaviour constraints of the various worker processes during their migration by means of constraint orchestration. We do this for the above migration scenario only. Thus, Figure 12 presents main migration coordination details, in terms of the *Evol* partitions for the one scenario.

The figure has been annotated with seven notes: $A, \ldots, G$. Each note indicates a relevant horizontal pictorial level: *McPal* performs a migrative step by progressively coordinating the simultaneous behavioural freedom of the original worker processes. Thus, at note $A$, *McPal* starts its own migrative phase only, like already specified in Figure 8. At note $B$, *McPal* simultaneously restricts the worker processes, by giving them a last orders possibility for their critical activity by means of phase *WorkerToBeUnit*. At note $C$, *McPal*, on the basis of its first observation of a trap *ready* of *WorkerToBeUnit* having been entered (by $Worker_2$), appoints $Worker_2$ as the future $Unit_1$. At note $D$, *McPal*, on the basis of its second observation of a trap *ready* of *WorkerToBeUnit* having been entered (by $Worker_1$), appoints $Worker_1$ as the future $Unit_2$ and moreover appoints $Worker_3$ as the future $Unit_4$. Note, the latter appointment is done on the basis of trap *triv* having been entered (instead of trap *ready*). At note $E$, on the basis of trap *onItsWay* of phase $TowardsUnit_4$ having been entered by $Worker_3$ as well as trap *triv* of phase *OrigAsSched* by *Scheduler*, *McPal* changes phase *OrigAsSched* of *Scheduler* into $TowardsUnit_3$. At note $F$, *McPal*, seeing the scheduler and each worker in their respective traps *ready* of the four phases $TowardsUnit_{..}$, simultaneously changes these phases into $ToBeAsUnit_{..}$. At note $G$, *McPal* finishes its own migrative phase, like already specified in Figure 8.

In exact conformity to the above constraint orchestration of the one scenario, the consistency rules below specify all six scenarios. Note how change clauses are used for parameterizing the actual scenario followed. We actually present seven rules, one for each note $A$ to $G$ and in the same order. We thereby repeat the rules for note $A$ and $G$, already presented as new rules in the context of Figure 8. As the seven rules specify migrative steps only, they all belong to the set $Crs_{migr}$. So, the last rule actually removes these seven rules, at the end of the migration indeed, together with the original rules no longer needed.

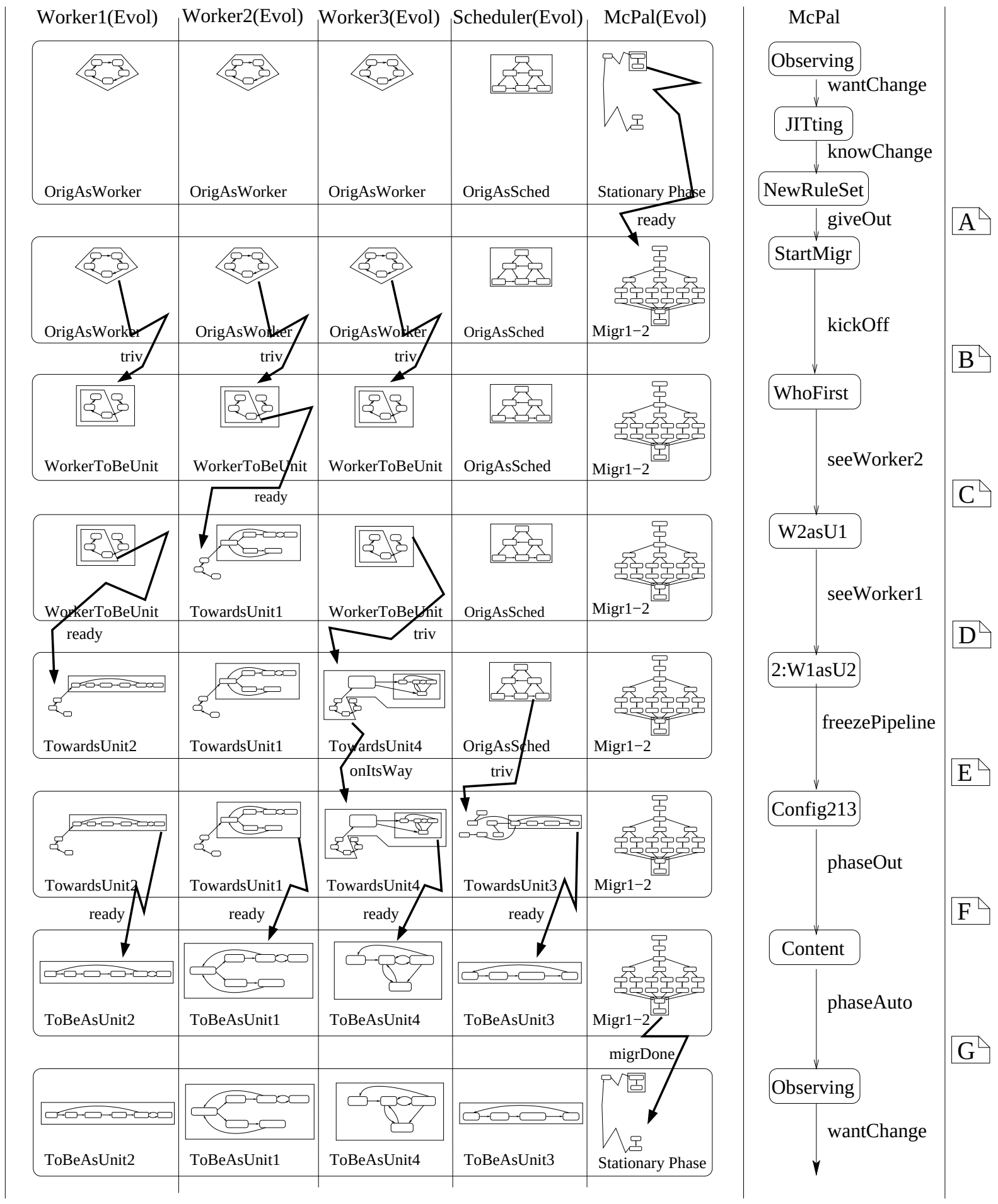

**Fig. 12.** Orchestrating a migration coordination scenario – main lines.

$$\mathit{McPal}\colon \mathit{NewRuleSet} \stackrel{\mathit{giveOut}}{\rightarrow} \mathit{StartMigr} \ * \ \mathit{McPal}(\mathit{Evol})\colon \mathit{StatPhase} \stackrel{\mathit{ready}}{\rightarrow} \mathit{Migr}_{1-2}$$

$$\begin{aligned}&\mathit{McPal}\colon \mathit{StartMigr} \stackrel{\mathit{kickOff}}{\rightarrow} \mathit{WhoFirst} \ * \\ &\quad \mathit{Worker}_1\,(\mathit{Evol})\colon \mathit{OrigAsWorker}_1 \stackrel{\mathit{triv}}{\rightarrow} \mathit{WorkerToBeUnit}, \\ &\quad \mathit{Worker}_2\,(\mathit{Evol})\colon \mathit{OrigAsWorker}_1 \stackrel{\mathit{triv}}{\rightarrow} \mathit{WorkerToBeUnit}, \\ &\quad \mathit{Worker}_3\,(\mathit{Evol})\colon \mathit{OrigAsWorker}_1 \stackrel{\mathit{triv}}{\rightarrow} \mathit{WorkerToBeUnit}\end{aligned}$$

$$\begin{aligned}&\mathit{McPal}\colon \mathit{WhoFirst} \stackrel{\mathit{seeWorker}_i}{\rightarrow} W_i\,\mathtt{as}\,U_1 \ * \\ &\quad \mathit{Worker}_i(\mathit{Evol})\colon \mathit{WorkerToBeUnit} \stackrel{\mathit{ready}}{\rightarrow} \mathit{TowardsUnit}_1\end{aligned}$$

$$\begin{aligned}&\mathit{McPal}\colon W_i\,\mathtt{as}\,U_1 \stackrel{\mathit{seeWorker}_j}{\rightarrow} i{:}W_j\,\mathtt{as}\,U_2 \ * \\ &\quad \mathit{Worker}_j\,(\mathit{Evol})\colon \mathit{WorkerToBeUnit} \stackrel{\mathit{ready}}{\rightarrow} \mathit{TowardsUnit}_2, \\ &\quad \mathit{Worker}_k\,(\mathit{Evol})\colon \mathit{WorkerToBeUnit} \stackrel{\mathit{triv}}{\rightarrow} \mathit{TowardsUnit}_4\end{aligned}$$

$$\begin{array}{l} McPal\colon i\colon W_j\,\mathtt{as}\,U_2 \stackrel{freezePipeline}{\rightarrow} Conf_{i,j,k} \;\; * \\ \quad Worker_k(Evol)\colon\; TowardsUnit_4 \stackrel{onItsWay}{\rightarrow} TowardsUnit_4, \\ \quad Scheduler\,(Evol)\colon OrigAsSched \stackrel{triv}{\rightarrow} TowardsUnit_3 \end{array}$$

$$\begin{array}{l} McPal\colon Conf_{i,j,k} \stackrel{phaseOut}{\rightarrow} Content \;\; * \\ \quad Worker_i\,(Evol)\colon TowardsUnit_1 \stackrel{ready}{\rightarrow} ToBeAsUnit_1, \\ \quad Worker_j\,(Evol)\colon TowardsUnit_2 \stackrel{ready}{\rightarrow} ToBeAsUnit_2, \\ \quad Scheduler\,(Evol)\colon TowardsUnit_3 \stackrel{ready}{\rightarrow} ToBeAsUnit_3, \\ \quad Worker_k\,(Evol)\colon TowardsUnit_4 \stackrel{ready}{\rightarrow} ToBeAsUnit_4 \end{array}$$

$$\begin{array}{l} McPal\colon Content \stackrel{phaseAuto}{\rightarrow} Observing \;\; * \\ \quad McPal\,(Evol)\colon Migr_{1-2} \stackrel{migrDone}{\rightarrow} StatPhase,\; McPal[\, Crs \;\mathtt{:=}\; Crs_{toBe}\,] \end{array}$$

As we have experienced, a constraint orchestration as presented is very helpful for understanding corresponding consistency rules. Furthermore, the collaboration snapshots illuminate *McPal* dynamics as well as understanding how to model, just-in-time, the phases and traps in the various *Evol* partitions and how to model, JIT too, the global processes at their levels. So, the various UML diagrams used: collaboration, state machine (simple and special, because of Paradigm) and activity diagram (special as constraint orchestration), do support the understanding of the dynamically consistent coordination of the originally unforeseen migration of a first architecture towards a second, JIT modeled architecture. As such migration is a special form of coordination, there is no temporary quiescence needed of whatever component: the migration is really on-the-fly. Moreover, after *McPal* has returned to its stand-by behaviours in its phase *StatPhase*, it is ready for starting yet another completely different and unforeseen migration in a dynamically consistent way.

## 4 Conclusion

To cater for future, unforeseen on-the-fly migration, *McPal* can be incorporated into any Paradigm model. *McPal* comprises the coordination needed for conducting the migration collaboration of all components towards a new way of working. Modeled just-in-time, migration coordination is such that no form of temporary quiescence is needed to occur for whatever component during its migration. Any such component remains in execution, be it in a smoothly changing manner, gradually adapting to new circumstances arising along the migration trajectory. *McPal*'s coordination, as specified in Paradigm, maintains consistency of the model both during and after migration, thereby dealing with the various dynamic consistency problem types. It is noted, UML visualizations substantially supplement the understanding of the evolution. In this manner, McPal is capable to coordinate its own migration trajectory dynamically consistent with the other components.

*McPal* as introduced here, generalizes the McPal from [16], as now it allows all kinds of complex behaviour in the JIT modeled phase *MigrPhase*, as long as it remains a correct Paradigm coordination. The McPal component proposed in [16] was restricted to a fixed migration coordination pattern. As a next and substantial generalization we are going to incorporate semantics for creating as well as deleting dynamics, in particular of type A and of type B, consistently. This will enable, e.g., creation of a stand-by *McPal* when the original *McPal* is busy with a first migration: the stand-by can then initiate a different migration, even before the first has finished. As Paradigm's processes can model both ICT activities and business activities, the McPal pattern is particularly promising for addressing all kinds of alignment situations between business and ICT and moreover for addressing general evolution of ICT and business in tandem, cf. [23].

By means of ParADE, a modeling and animation environment for Paradigm models, PhD work by Andries Stam, first migration examples have been implemented and tested. A preliminary, but promising result [1] has been obtained in translating Paradigm into the process algebra ACP. With its coupling to the mCRL2 toolkit [18], state-of-the-art modelchecking of Paradigm models, in particular of migration trajectories, comes into reach.

## 5 Appendix

The appendix lists the definitions of Paradigm's main syntactical ingredients, viz. process, phase, (connecting) trap, partition and global process, and very briefly indicates their meanings. It ends with presenting Paradigm's consistency rules on which Paradigm's operational semantics is based.

**Definition 1.**

*A* process *or* state-transition diagram *or* STD $Z$ *is a triple* $\langle \mathtt{ST}, \mathtt{AC}, \mathtt{TS} \rangle$*. Here,* $\mathtt{ST}$ *is the non-empty set of* states*,* $\mathtt{AC}$ *is the set of* actions *or* transition labels *and* $\mathtt{TS} \subseteq \mathtt{ST} \times \mathtt{AC} \times \mathtt{ST}$ *is the set of* transitions*. A transition* $(x, a, x') \in \mathtt{TS}$ *is said to be* from $x$ to $x'$ *and is denoted as* $x \xrightarrow{a} x'$*.*

*A* phase *or* subprocess $S$ *of an STD* $Z = \langle \mathtt{ST}, \mathtt{AC}, \mathtt{TS} \rangle$ *is itself an STD* $S = \langle \mathtt{st}, \mathtt{ac}, \mathtt{ts} \rangle$ *with* $\mathtt{st} \subseteq \mathtt{ST}$*,* $\mathtt{ac} \subseteq \mathtt{AC}$*,* $\mathtt{ts} \subseteq \{ (x, a, x') \in \mathtt{TS} \mid x, x' \in \mathtt{st}, a \in \mathtt{ac} \}$*.*

*A* trap $t$ *of a phase* $S = \langle \mathtt{st}, \mathtt{ac}, \mathtt{ts} \rangle$ *is a nonempty set of states* $t \subseteq \mathtt{st}$ *such that* $x \in t$ *and* $x \xrightarrow{a} x' \in \mathtt{ts}$ *imply that* $x' \in t$*. If* $t = \mathtt{st}$*, the trap is called* trivial*, denoted as triv or triv(*$S$*).*

*For two phases* $S = \langle \mathtt{st}, \mathtt{ac}, \mathtt{ts} \rangle$ *and* $S' = \langle \mathtt{st}', \mathtt{ac}', \mathtt{ts}' \rangle$ *be of the same STD, a trap* $t$ *of* $S$ *is called* connecting from $S$ to $S'$ *if* $t \subseteq \mathtt{st}'$*. The triple* $(S, t, S')$ *is then called a* phase change*, notation* $S \xrightarrow{t} S'$*.*

*A* partition $\{ (S_i, T_i) \mid i \in I \}$ *of an STD* $Z = \langle \mathtt{ST}, \mathtt{AC}, \mathtt{TS} \rangle$ *is a nonempty set of phases* $S_i = \langle \mathtt{st}_i, \mathtt{ac}_i, \mathtt{ts}_i \rangle$ *of* $Z$*, each with a set* $T_i$ *of its traps with triv(*$S_i$*)* $\in T_i$*. If* $\mathtt{ST} = \bigcup_{i \in I} \mathtt{st}_i$ *and* $\mathtt{TS} = \bigcup_{i \in I} \mathtt{ts}_i$*, the partition is said to be* covering *process* $Z$ *or* covering *for short.*

*A* global process at the level of *a partition* $\pi = \{ (S_i, T_i) \mid i \in I \}$ *of another STD* $Z = \langle \mathtt{ST}, \mathtt{AC}, \mathtt{TS} \rangle$ *is an STD* $Z(\pi) = \langle \mathtt{GST}, \mathtt{GAC}, \mathtt{GTS} \rangle$ *with phases* $\mathtt{GST} \subseteq \{ S_i \mid i \in I \}$ *as its states, traps* $\mathtt{GAC} \subseteq \bigcup_{i \in I} T_i$ *as its actions, and phase changes* $\mathtt{GTS} \subseteq \{ S_i \xrightarrow{t} S_j \mid i, j \in I, t \in \mathtt{GAC} \}$ *as its transitions. The STD* $Z$ *is called* underlying *the global STD* $Z(\pi)$ *or* more detailed *than* $Z(\pi)$ *or just* detailed*.*

A process presents all (sequential) dynamics a certain unit (component, thread, object, person, team, etc.) can have or can undergo as far as relevant in a certain situation. A phase of a process is a temporary constraint imposed on the process, restricting the process to the part expressed as phase. A trap of a phase is a temporary constraint committed to by that phase, deliberatively taken up by entering the set of states specified as trap. A global process is a process specifying the (sequential) constraint dynamics of the underlying detailed process as far as relevant to a certain (collaborative) situation. As such, a global process has as states a set of phases and has as actions connecting traps from one phase (as global state) to another phase (as next global state). One underlying, detailed process can have many global processes, each defined in terms of separated sets of phases and their traps. Such a separate set is determined by a partition.

In 'normal', foreseen coordination situation, any partition is covering; otherwise, what were not covered by it, would forever be excluded from occurring or being reached. But, in (originally) unforeseen coordination situations, arising when lazily modelled migration is to be coordinated, it should be possible to extend partitions later. So their being covering is no longer needed, rather counterproductive, as later modeled constraint dynamics can allow as yet what before was excluded from occurring or from being reached.

**Definition 2.** *Let* $\{\ P_i \mid i \in I\ \}$ *be a set of detailed processes. For each* $i \in I$*, let* $J_i$ *be another index set with* $0 \in J_i$ *and let* $\{\ \pi_{i,j} \mid j \in J_i, j \neq 0\ \}$ *be a set of partitions of process* $P_i$*. The product space of the state spaces of these processes is called the* Cartesian frame *of these processes, denoted as CF.*

*A* consistency rule *cr for CF has, in its complete form, the following format*

$$P_m\colon s \xrightarrow{a} s' \ \ * \ \ P_e(\pi_{e,p})\colon S_{e,p} \xrightarrow{\theta_{e,p}} S'_{e,p},\ \ldots,\ P_f(\pi_{f,q})\colon S_{f,q} \xrightarrow{\theta_{f,q}} S'_{f,q},\ \ P_m[\boldsymbol{x}\colon= \boldsymbol{\alpha}]$$

*where* $\boldsymbol{x}$ *is a sequence of variables local to the detailed process* $P_m$ *and* $\boldsymbol{\alpha}$ *is a corresponding sequence of (expressions of) values.*

The part before the $*$ is called the *detailed step* of the consistency rule *cr*; it contains either zero or one detailed process transition. The part right of the $*$ of the processes $P_e$ to $P_f$ is called the *protocol step*; it contains zero or more phase changes, each from a different global process in frame *CF*. The right-most part, recognizable from its square brackets involving process $P_m$, is called the *change clause*; it contains one (or more) assignments to variables of the lefthand process $P_m$. Optionally, the part left of $*$, process name and transition, can be omitted. In that case, the part with a change clause is absent too.

A consistency rule *cr*, via the coupling in its format, specifies a (possible) synchronization of single steps taken simultaneously in different coordinates of the Cartesian frame *CF*. If the protocol step is non-empty, detailed process $P_m$ is called *manager* of its *employees* being the different detailed processes $P_e, \ldots, P_f$. Such synchonized step in the frame *CF* can be taken only, if not only the detailed step is contained in each current phase of detailed process $P_m$ but also, within each current phase $S_{e,p}, \ldots, S_{f,q}$, the connecting traps $\theta_{e,p}, \ldots, \theta_{f,q}$ have been entered, respectively.

A consistency rule specifies which phase changes can occur simultaneously, strictly synchronized and at most one per partition / global process. Such synchronized ensemble of global steps can be additionally synchronized with at most one detailed step and if so, the larger synchronized ensemble can be even further synchonized with updates of one or more variables local to the same detailed process. Please note, constraints imposed as well as constraints committed to really restrict the taking of the detailed step as well as the taking of the protocol step of consistency rule *cr*.